\documentclass[a4paper,11pt]{article}
\pdfoutput=1
\usepackage{jheppub} 
\usepackage[utf8]{inputenc}
\usepackage[T1]{fontenc} 
\usepackage{slashed}
\usepackage{xspace}
\usepackage{dsfont}

\title{\boldmath Extending the Universal One-Loop Effective Action by
  Regularization Scheme Translating Operators}

\author{Benjamin Summ}
\author{and Alexander Voigt}
\affiliation{Institute for Theoretical Particle Physics and Cosmology,\\
  RWTH Aachen University, 52074 Aachen, Germany}
\emailAdd{benjamin.summ@rwth-aachen.de}
\emailAdd{alexander.voigt@physik.rwth-aachen.de}

\abstract{We extend the universal one-loop effective action (UOLEA) by
  operators which translate between dimensional reduction (DRED) and
  dimensional regularization (DREG).  These regularization scheme
  translating operators allow for an application of the UOLEA to
  supersymmetric high-scale models matched to non-supersymmetric
  effective theories.  The operators are presented in a generic, model
  independent form, suitable for implementation into generic spectrum
  generators.}


\hypersetup{
  pdftitle    = {Extending the Universal One-Loop Effective Action by Regularization Scheme Translating Operators},
  pdfauthor   = {Benjamin Summ, Alexander Voigt},
  pdfkeywords = {UOLEA, DREG, DRED}
}

\newcommand{\MSbar}{\ensuremath{\overline{\text{MS}}}\xspace}
\newcommand{\DRbar}{\ensuremath{\overline{\text{DR}}}\xspace}
\newcommand{\DRbarPrime}{\ensuremath{\overline{\text{DR}}'}\xspace}
\newcommand{\DRED}{\ensuremath{\text{DRED}}\xspace}
\newcommand{\DREG}{\ensuremath{\text{DREG}}\xspace}
\newcommand{\GeV}{\ensuremath{\;\text{GeV}}}
\newcommand{\FS}{\texttt{FlexibleSUSY}\@\xspace}
\newcommand{\FeynRules}{\texttt{FeynRules}\@\xspace}
\newcommand{\SARAH}{\texttt{SARAH}\@\xspace}
\newcommand{\Lag}{\mathcal{L}}
\newcommand{\Lagint}{\mathcal{L}_{\text{int}}}

\newcommand{\gc}{\ensuremath{g_{\lambda}}}
\newcommand{\A}{\mathcal{A}}
\newcommand{\B}{\mathcal{B}}
\DeclareMathOperator{\tr}{tr}

\begin{document} 

\begin{flushright}
\footnotesize
TTK--18--19
\end{flushright}

\maketitle
\flushbottom

\section{Introduction}
\label{sec:introduction}

With the discovery of the Higgs boson at the Large Hadron Collider
(LHC) \cite{Aad:2012tfa,Chatrchyan:2012xdj} it became clear that the
Standard Model of Particle Physics (SM) is a good description of the
physics at and below the electroweak scale.  However, it is also clear that
the SM does not provide a complete description of nature, as it fails
to describe phenomena such as dark matter and does not incorporate
gravity, for example.  Besides these weaknesses, there are also many
open questions, e.g., whether the electroweak vacuum is stable up to
the Planck scale or whether there is a hierarchy problem and how it
may be avoided.  Many models beyond the SM (BSM) have been proposed to
address the open questions and the drawbacks of the SM.  One of the
promising SM extensions is supersymmetry (SUSY), which can provide a
solution to the hierarchy problem, explain the deviation of the
anomalous magnetic moment of the muon and the stability of the
electroweak vacuum.  However, no supersymmetric particles with masses
below the TeV scale have been discovered so far, which means that if
supersymmetry is realized in nature, the SUSY particles may be heavier
than the TeV scale.  This finding is supported by the measured value
of the Higgs boson mass of $M_h = 125.09\pm 0.32\GeV$ \cite{Aad:2015zhl}: SUSY
models often predict the mass of the SM-like Higgs boson to be of the order
of the $Z$ boson mass, $M_Z = 91.1876\GeV$, at tree-level.  In order
to raise the predicted Higgs mass to its measured value, large loop
corrections are required, which can be achieved by the presence of
multi-TeV colored SUSY particles.  Large loop corrections, on the
other hand, spoil the convergence of the perturbation series, leading
to large uncertainties in fixed-order calculations.

Effective field theories (EFTs) are a well suited approach to obtain
precise low-energy predictions of BSM models with heavy particles.
However, depending on the mass hierarchy of the studied high-scale
model, many different EFTs must be considered.  In order to avoid
repetition in the derivation of all possible EFT Lagrangians,
the universal one-loop effective action (UOLEA) has been developed
\cite{Henning:2014wua,Drozd:2015rsp,Ellis:2017jns}.  It
provides generic expressions for the Wilson coefficients of the
operators of the effective Lagrangian up to 1-loop level and dimension
six.  These generic expressions are well suited to be implemented into
generic spectrum generators such as \FeynRules
\cite{Christensen:2008py,Duhr:2011se,Alloul:2013fw,Alloul:2013bka},
\FS \cite{Athron:2014yba,Athron:2017fvs} or \SARAH
\cite{Staub:2009bi,Staub:2010jh,Staub:2012pb, Staub:2013tta} to
calculate precise predictions in all possible low-energy EFTs in a
fully automated way.

The currently known effective operators of the UOLEA
\cite{Henning:2014wua,Drozd:2015rsp,Ellis:2017jns} are renormalized in
the \MSbar scheme. Although this scheme is well suited to
renormalize non-supersymmetric models, it is cumbersome to apply it to
supersymmetric models, because the underlying dimensional
regularization (DREG) \cite{tHooft:1972tcz} explicitly breaks
supersymmetry \cite{Delbourgo:1974az}.  To nevertheless perform loop calculations
in an \MSbar renormalized SUSY model one would have to restore
supersymmetry, for example by introducing supersymmetry-restoring
counter terms, as discussed for example in
\cite{Martin:1993yx,Mihaila:2009bn,Stockinger:2011gp}.  In
supersymmetric models
regularization by dimensional reduction (DRED) \cite{Siegel:1979wq} is
currently known to not break supersymmetry up to the 3-loop level
\cite{Capper:1979ns,Stockinger:2005gx,Stockinger:2018oxe}
and is therefore widely adopted in SUSY loop calculations.
In order to apply the UOLEA to a scenario, where heavy particles of a
supersymmetric model (renormalized in the \DRbar scheme) are
integrated out at a high scale and a non-supersymmetric EFT
(renormalized in the \MSbar scheme) results at low energies, the
change of the regularization scheme from DRED to DREG must be
accounted for by shifting the running parameters by finite terms.
For general renormalizable softly broken supersymmetric gauge theories these
parameter shifts have been known at the 1-loop level
for a long time \cite{Martin:1993yx}.  However, in the formalism of
the UOLEA the generic effective operators that correspond to such a
regularization scheme change are currently unknown and reconstructing
them from the results of Ref.~\cite{Martin:1993yx} is difficult
due to the presence of finite field renormalizations.

In this paper we present all 1-loop effective operators that appear in
the effective Lagrangian when changing the regularization scheme from
DRED to DREG, assuming that the (not necessarily supersymmetric) UV model
is renormalizable.  We perform the calculation
in the formalism of effective field theories by making use of the fact
that the difference between DRED and DREG can be expressed by the
presence/absence of so-called $\epsilon$-scalars \cite{Capper:1979ns}.
The $\epsilon$-scalars are integrated out from the DRED-regularized UV model and the
resulting effective operators are formulated in the language of the
UOLEA.  Our generic results complement the currently known
generic expressions of the UOLEA and allow for its application to
supersymmetric high-scale models and its implementation into generic
spectrum generators.  Finally, we show that our results are in agreement
with the known generic parameter conversion terms of
Ref.~\cite{Martin:1993yx}.

In Section~\ref{sec:epsilon} we briefly review the formalism of
$\epsilon$-scalars in DRED and give projection relations and Lagrangian terms
necessary for the calculation of the regularization scheme translating
operators, which we derive in Section~\ref{sec:results}.  We apply our
derived effective Lagrangian in Section~\ref{sec:applications} to the
general supersymmetric model of Ref.~\cite{Martin:1993yx} to show that
our results reproduce the parameter relations derived in that
reference.  We conclude in Section~\ref{sec:conclusion}.

\section{Epsilon scalars in dimensional reduction}
\label{sec:epsilon}

In the following we briefly review the relation between DRED and DREG,
relevant to the derivation of the effective Lagrangian in
Section~\ref{sec:results}.  In DRED an infinite dimensional space is
introduced, which has the characteristics of a $4$-dimensional space,
denoted as $Q4S$. This quasi-$4$-dimensional space is decomposed
as $Q4S=QdS\oplus Q\epsilon S$, where $QdS$ is an infinite dimensional
space that is formally $d$-dimensional and $Q\epsilon S$ is its
complement, formally of dimension $\epsilon=4-d$
\cite{Stockinger:2005gx}. The metrics of the spaces $Q4S$, $QdS$ and
$Q\epsilon S$ are denoted by $g^\mu_\nu$, $\hat{g}^\mu_\nu$ and
$\tilde{g}^\mu_\nu$, respectively, and satisfy
\begin{align}
\label{contraction1}
g^\mu_\nu&=\hat{g}^\mu_\nu+\tilde{g}^\mu_\nu\text{,} & g^\mu_\mu=4\text{,} \\
\label{contraction2}
g^{\mu\nu}\tilde{g}^\rho_\nu&=\tilde{g}^{\mu\rho}\text{,} & \tilde{g}^\mu_\mu=\epsilon\text{,} \\
\label{contraction3}
g^{\mu\nu}\hat{g}^\rho_\nu&=\hat{g}^{\mu\rho}\text{,} & \hat{g}^\mu_\mu=d\text{,}\\
\hat{g}^{\mu\nu}\tilde{g}^\rho_\nu&=0\text{.}
\end{align}
The signature of the metric of $Q\epsilon S$ is $(-,-,\dots)$. In DRED
momenta are taken to be $d$-dimensional, whereas gauge fields and
$\gamma$-matrices are taken to be $4$-dimensional. We use the
convention of a totally anti-commuting $\gamma^5$. Due to the
decomposition of $Q4S$ it is convenient to split the gauge field
$A^a_\mu\in Q4S$ into two parts,
$A^a_\mu = \hat{A} ^a _\mu + \epsilon ^a _\mu$, with
$\hat{A}^a_\mu \in QdS$ and $\epsilon ^a_\mu \in Q\epsilon S$. The
$\epsilon$-dimensional field $\epsilon ^a_\mu$ is a scalar under
$d$-dimensional Lorentz transformations and is referred to as
$\epsilon$-scalar \cite{Capper:1979ns}.  With respect to the gauge
group associated with $A^a_\mu$ the $\epsilon$-scalar transforms in
the adjoint representation.  After the gauge field has been split in
this way, the Lagrangian may contain the following additional terms
with $\epsilon$-scalars,
\begin{align}
  \Lag &= \Lag_\phi + \Lag_\psi + \Lag_\epsilon,
  \label{eq:Lag_general_epsion} \\
  \Lag_\phi &= \epsilon_\mu^a\epsilon^\mu_b F^a_b[\phi_1,\phi_2,\ldots \phi_n],
  \label{epsilonscalarcoupling} \\
  \Lag_\psi &= \epsilon_\mu^a \bar{\psi}_i \tilde{\gamma}^\mu \Gamma T^a_{ij}\psi_j,
  \label{epsilonfermioncoupling} \\
  \Lag_\epsilon &= -\frac{1}{2}(D^\mu\epsilon_\nu)^a (D_\mu \epsilon^\nu)_a+\frac{1}{2}m_\epsilon^2 \epsilon_\mu^a \epsilon^\mu_a-\frac{1}{4}g^2f^{abc}f^{ade}\epsilon^\mu_b \epsilon_\mu^d\epsilon^\nu_c\epsilon_\nu^e,
  \label{epsilonvectorcoupling}
\end{align}
where $\phi_i$ and $\psi_i$ denote scalars and fermions, respectively.
In Eq.~\eqref{epsilonscalarcoupling} $F^a_b$ is a function of the
scalar fields and may contain linear and quadratic terms. The symbol
$\tilde{\gamma}^\mu$ denotes a $\gamma$-matrix projected onto
$Q\epsilon S$, $\tilde{\gamma}^\mu = \tilde{g}^{\mu}_\nu \gamma^\nu$,
and $\Gamma$ is some $4\times 4$ matrix that contains products of
$\{\mathbf{1}, \gamma^\mu, \gamma^5 \}$.  In the following we denote
any projection of a Lorentz tensor $T^{\sigma \rho \cdots}$ onto
$Q\epsilon S$ by
$\tilde{T}^{\mu \nu \cdots}=\tilde{g}^\mu_\sigma
\tilde{g}^\nu_\rho\cdots T^{\sigma \rho \cdots}$.  Similarly, tensors
projected onto $QdS$ are denoted as $\hat{T}^{\mu\nu\dots}$.
The $m^2_\epsilon$-dependent term in Eq.~\eqref{epsilonvectorcoupling}
can be removed by shifting the mass terms of the scalar fields
$\phi_i$ as described in Ref.~\cite{Jack:1994rk}, i.e.\ by changing
the renormalization scheme from \DRbar to \DRbarPrime.
Nevertheless, due to the remaining extra $\epsilon_\mu^a$-dependent
terms in the Lagrangian
\eqref{eq:Lag_general_epsion}, the difference between DRED and DREG
manifests in the presence of extra Feynman diagrams with
$\epsilon$-scalars, which contribute additional finite terms to
divergent loop amplitudes due to the contraction relation
\eqref{contraction2}.

In the following section we integrate out the $\epsilon$-scalars using
the language of effective field theories. In the limit
$\epsilon\rightarrow 0$ this effectively results in a change of the
regularization scheme from DRED to DREG.  The resulting additional
finite 1-loop operators that appear in the ``effective'' Lagrangian
can be absorbed by a re-definition of the fields and the running
parameters, leading to the same parameter relations given in
Ref.~\cite{Martin:1993yx}.

\section{Regularization scheme translating operators in the UOLEA}
\label{sec:results}

To derive the operators that translate between DRED and DREG we
consider a general renormalizable gauge theory with the gauge
group $G$ and the Lagrangian $\Lag$, which contains real scalar fields
$\phi_i$, Dirac fermions $\psi_i$ and a set of four-component Majorana
fermions $\lambda_i$.\footnote{The formulation of the Lagrangian in
  terms of Dirac and Majorana fermions has been chosen in order to
  diagonalize the operator
  $(i\hat{\slashed{D}}-m)$.} We
furthermore assume that the theory is regularized in DRED.  The gauge
field $A^a_\mu$ is split into a $d$- and an $\epsilon$-dimensional
component, as described in Section~\ref{sec:epsilon}, and we
distinguish the $\epsilon$-scalars from the scalars $\phi_i$.

To calculate the effective action up to the 1-loop level, we first
split all fields
$\omega_i \in \{\phi_i, \psi_i, \lambda_i, \hat{A}^a_\mu,
\epsilon^a_\mu \}$ into a background part $\omega_{B,i}$, satisfying
the classical equations of motion, and a corresponding fluctuation
$\delta\omega_i$. The calculation is going to be performed using a
covariant derivative expansion
\cite{Gaillard:1985uh,CHEYETTE1988183,Henning:2014wua} in order to
obtain a manifestly gauge invariant result. This means in particular
that the operator\footnote{Note that whereas this notation suggests that we are treating a simple gauge group we are not restricted to this case. The notation is to be understood with a sum over all factors of the gauge group with their respective gauge couplings.}
$\hat{P}_\mu\equiv i\hat{D}_\mu=i\hat{\partial}_\mu+g\hat{A}^a_{B,\mu}
T^a$, where
$\hat{A}^a_{B,\mu}$ is the background gauge field, should be kept as a
whole in the calculation and not be split into $\hat{\partial}_\mu$
and $\hat{A}^a_{B,\mu}$. Furthermore, to obtain an action which is
gauge invariant under transformations of $\hat{A}^a_{B,\mu}$ we only
fix the gauge of the fluctuation $\delta \hat{A}^a_\mu$
\cite{Abbott:1981ke}. We choose a gauge fixing Lagrangian of the
form \cite{Henning:2014wua}
\begin{align}
\label{eq:gaugefixing}
\Lag _{\text{g.f.}}=-\frac{1}{2 \xi}\left[\xi (m_A)_{ab} \eta^b+\hat{D}^\mu\delta \hat{A}^a_\mu\right]^2,
\end{align}
where the fields $\eta^a$ are the Goldstone bosons corresponding to
the spontaneously broken generators of the gauge group and $m_A$ is
the diagonal mass matrix of the gauge bosons. The part of the
Lagrangian containing the ghost fields is given by
\begin{align}
\label{ghosts}
\Lag_\text{ghost}=\bar{c}^a(-\hat{D}^2-\xi m_A^2)_{ab}c^b \text{.}
\end{align}
In the following the Goldstone bosons are not treated separately, but
are regarded as part of the vector of scalar fields
$\phi_i$. Moreover, for the purpose of this calculation the
fluctuation $\delta \hat{A}^a_\mu$ can be treated as a scalar field
transforming in the adjoint representation under background gauge
transformations \cite{Henning:2014wua}. Similarly, the ghosts can be
regarded as usual fermions in the adjoint representation of the gauge
group.  In the following calculation the path integral over the ghosts
can be performed directly and is independent of $\epsilon$-scalars.
The ghosts will therefore not be considered further in this paper.
The second variation of the action around the background fields then
reads
\begin{align}
\begin{split}
\delta^2\mathcal{L}&=\delta\bar{\psi}\Delta_{\psi}\delta\psi+\delta\bar{\lambda}\Delta_{\lambda}\delta\lambda-\frac{1}{2}\delta\epsilon_\mu\tilde{\Delta}^{\mu\nu}_{\epsilon}
\delta\epsilon_\nu-\frac{1}{2}\delta\Phi\Delta_{\Phi}\delta\Phi \\
 &\quad-\delta\bar{\psi}\tilde{X}^\mu_{\bar{\psi}\epsilon}\delta\epsilon_\mu
-\delta\bar{\psi}X_{\bar{\psi}\Phi}\delta\Phi-\delta\bar{\lambda}\tilde{X}^\mu_{\bar{\lambda}\epsilon}\delta\epsilon_\mu
-\delta\bar{\lambda}X_{\bar{\lambda}\phi}\delta\Phi \\
 &\quad +\delta\epsilon_\mu \tilde{X}^\mu_{\epsilon\psi}\delta\psi+\delta\Phi X_{\Phi\psi}\delta\psi+\delta\epsilon_\mu \tilde{X}^\mu_{\epsilon\lambda}\delta\lambda
+\delta\Phi X_{\Phi\lambda}\delta\lambda \\
& \quad +\delta\bar{\psi}X_{\bar{\psi}\lambda}\delta\lambda+\delta\bar{\lambda}X_{\bar{\lambda}\psi}\delta\psi-\frac{1}{2}\delta\epsilon_\mu \tilde{X}^\mu_{\epsilon\Phi}\delta\Phi-\frac{1}{2}\delta\Phi \tilde{X}^\mu_{\Phi\epsilon}\delta\epsilon_\mu\text{,}
\end{split}
\label{fullsecondvar}
\end{align}
where
\begin{align}
\delta \Phi&=\begin{pmatrix}
\delta \phi \\ \delta \hat{A}_\mu
\end{pmatrix},
\\
X_{\Phi \omega}&=\begin{pmatrix}
X_{\phi \omega} \\ \hat{X}^\mu_{\hat{A} \omega}
\end{pmatrix},
\\
X_{\omega \Phi}&=\begin{pmatrix}
X_{\omega \phi} &\hat{X}^\mu_{\omega \hat{A}}
\end{pmatrix},
\end{align}
and
\begin{align}
  X_{\omega \sigma}\equiv-\left.\frac{\delta^2\Lagint}{\delta \omega \delta \sigma}\right\rvert
\end{align}
denotes the derivative of the interaction Lagrangian, $\Lagint$, with
respect to the fields $\omega$ and $\sigma$, evaluated at the
background field configuration.  Furthermore we have introduced the
abbreviations
\begin{align}
  \Delta_{\Phi}&\equiv\begin{pmatrix}
  \Delta_\phi & \hat{X}^\mu_{\phi \hat{A}} \\
  \hat{X}^\mu_{\hat{A} \phi} & \hat{\Delta}_{\hat{A}}^{\mu \nu}
  \end{pmatrix}, \\
  \Delta_{\psi}&\equiv\slashed{P}-m_\psi+X_{\bar{\psi}\psi}\text{,} \\
  \Delta_{\lambda}&\equiv\frac{1}{2}\slashed{P}-\frac{1}{2}m_\lambda+X_{\bar{\lambda}\lambda}\text{,} \\
  \Delta_{\phi}&\equiv-P^2+m_\phi^2+X_{\phi\phi}\text{,}\\
  \label{Eq:Vectorfieldprop}
  \hat{\Delta}^{\mu \nu}_{\hat{A}}&\equiv  P^2\hat{g}^{\mu \nu}-2\hat{P}^\nu \hat{P}^\mu+\hat{P}^\mu \hat{P}^\nu \left(1+\frac{1}{\xi}\right)+m_A^2 \hat{g}^{\mu \nu},\\
  \tilde{\Delta}^{\mu \nu}_{\epsilon}&\equiv\tilde{g}^{\mu\nu} (P^2-m_\epsilon^2)+\tilde{X}^{\mu\nu}_{\epsilon \epsilon}\text{.}
\end{align}
In any product that contains $\Phi$ the Lorentz indices are fully
contracted, for example
\begin{align}
\delta \Phi X_{\Phi \psi} \delta \psi= \begin{pmatrix}
\delta \phi & \delta \hat{A}_\mu 
\end{pmatrix}  
\begin{pmatrix}
X_{\phi \psi} \\ \hat{X}^\mu_{\hat{A}\psi}
\end{pmatrix}
\delta \psi\text{.}
\end{align}
In addition, in Eq.~\eqref{fullsecondvar} all indices, except for the
Lorentz indices of $Q\epsilon S$, have been suppressed for
brevity. Eq.~\eqref{fullsecondvar} can be simplified further due to
the constraints on the possible couplings of $\epsilon$-scalars to
other fields as given in
Eqs.~\eqref{eq:Lag_general_epsion}--\eqref{epsilonvectorcoupling}: We
can solve the classical equations of motion in a perturbation
expansion in couplings. The leading term is proportional to an
operator of the form $\bar{\psi}\tilde{\gamma}^\mu\psi$ and thus every
term in the series will contain this operator. In the limit
$\epsilon \to 0$ this operator vanishes, which means that the
background fields of the $\epsilon$-scalars can be set to zero from
the start.  This property can be used to simplify
Eq.~\eqref{fullsecondvar}, because from
Eqs.~\eqref{epsilonscalarcoupling} and \eqref{epsilonvectorcoupling}
it follows that
$\tilde{X}^\mu_{\Phi \epsilon}=\tilde{X}^\mu_{\epsilon \Phi}=0$ for
vanishing $\epsilon$-scalar background fields.

To perform the path integral, we shift the Dirac and Majorana fermions
to eliminate terms with mixed fermionic and bosonic fluctuations as
described in Ref.~\cite{Haba:2011vi}.  We first shift the Majorana
fermions by
\begin{align}
  \label{shift1}
  \delta\lambda'&=\delta\lambda-\Delta_\lambda^{-1}\left[\tilde{X}^\nu_{\bar{\lambda}\epsilon}
\delta\epsilon _\nu+X_{\bar{\lambda}\Phi}\delta\Phi-X_{\bar{\lambda}\psi}\delta\psi\right]\text{,} \\
  \label{shift2}
\delta\bar{\lambda}'&=\delta\bar{\lambda}+\left[\delta\epsilon _\mu \tilde{X}^\mu_{\epsilon\lambda}+\delta\Phi X_{\Phi\lambda}+\delta\bar{\psi}X_{\bar{\psi}\lambda}\right]\Delta_\lambda^{-1}, \\
  \intertext{and afterwards the Dirac fermions by}
  \label{shift3}
  \delta\psi'&=\delta\psi-\Lambda_\psi^{-1}\left[\tilde{\Xi}^\nu_{\bar{\psi}\epsilon}
\delta\epsilon _\nu+\Xi_{\bar{\psi}\Phi}\delta\Phi\right]\text{,} \\
  \label{shift4}
  \delta\bar{\psi}'&=\delta\bar{\psi}+\left[\delta\epsilon _\mu \tilde{\Xi}^\mu_{\epsilon\psi}+\delta\Phi \Xi_{\Phi\psi}\right]\Lambda_\psi^{-1} ,
\end{align}
and introduce the following abbreviations
\begin{align}
\tilde{\Lambda}^{\mu \nu}_\epsilon&=\tilde{\Delta}^{\mu\nu}_{\epsilon}-2\tilde{X}^\mu_{\epsilon\lambda}\Delta_\lambda^{-1}\tilde{X}^\nu_{\bar{\lambda}\epsilon}, \\
\Lambda_\Phi &=\Delta_{\Phi}-2X _{\Phi\lambda}\Delta_{\lambda}^{-1}X_{\bar{\lambda}\Phi}, \\
\Lambda_\psi&=\Delta_{\psi}-X_{\bar{\psi}\lambda}\Delta_\lambda^{-1}X_{\bar{\lambda}\psi}, \\
 \tilde{\Xi}^\mu_{\bar{\psi}\epsilon}&=\tilde{X}^\mu_{\bar{\psi}\epsilon}-X_{\bar{\psi}\lambda}\Delta_{\lambda}^{-1}\tilde{X}^\mu_{\bar{\lambda}\epsilon},\\
\Xi_{\bar{\psi} \Phi}&=X_{\bar{\psi}\Phi}-X_{\bar{\psi}\lambda}\Delta_{\lambda}^{-1}X_{\bar{\lambda}\Phi},\\
\tilde{\Xi}^\mu_{\epsilon \psi}&= \tilde{X}^\mu_{\epsilon\psi}-\tilde{X}^\mu_{\epsilon\lambda}\Delta_\lambda^{-1}X_{\bar{\lambda}\psi}, \\
\Xi_{\Phi \psi}&=X_{\Phi\psi}-X_{\Phi\lambda}\Delta_\lambda^{-1}X_{\bar{\lambda}\psi}, \\
 \tilde{\Xi}^\mu_{\epsilon\Phi} &= -2\tilde{X}^\mu_{\epsilon\lambda}\Delta_\lambda^{-1}X_{\bar{\lambda}\Phi} , \\
 \tilde{\Xi}^\mu_{\Phi\epsilon} &= -2X_{\Phi\lambda}\Delta_\lambda^{-1}\tilde{X}^\mu_{\bar{\lambda}\epsilon} .
\end{align}
For Dirac fermions the shifts \eqref{shift3}--\eqref{shift4} can be
performed independently. For Majorana fermions $\lambda$ and
$\bar{\lambda}$ are not independent and it is necessary that
$\delta\lambda' \gamma^0=\delta\bar{\lambda}'$ for the shifted fields
$\delta\lambda'$ and $\delta\bar{\lambda}'$ defined in
\eqref{shift1}--\eqref{shift2}, respectively. That this is indeed the
case is shown in Appendix \ref{sec:Appendix}.  After shifting the
fermions in this way, the variation takes the form
\begin{align}
\begin{split}
\delta^2\mathcal{L}&=\delta\bar{\psi}'\Lambda_\psi\delta\psi'+\delta\bar{\lambda}'\Delta_{\lambda}\delta\lambda'
-\frac{1}{2}\delta\Phi(\Lambda_{\Phi}-2\Xi_{\Phi \psi}\Lambda^{-1}_{\psi}\Xi_{ \bar{\psi}\Phi})\delta\Phi \\
 &\quad-\frac{1}{2}\delta\epsilon_\mu(\tilde{\Lambda}^{\mu\nu}_{\epsilon}-2\tilde{\Xi}^\mu_{\epsilon \psi}\Lambda^{-1}_{\psi}\tilde{\Xi}^\nu_{ \bar{\psi}\epsilon})
\delta\epsilon_\nu  
  -\frac{1}{2}\delta\epsilon_\mu (\tilde{\Xi}^\mu_{\epsilon\Phi}-2\tilde{\Xi}^\mu_{\epsilon \psi}\Lambda^{-1}_{\psi}\Xi_{ \bar{\psi}\Phi})\delta\Phi \\
&\quad-\frac{1}{2}\delta\Phi (\tilde{\Xi}^\mu_{\Phi\epsilon}-2\Xi_{\Phi \psi}\Lambda^{-1}_{\psi}\tilde{\Xi}^\mu_{ \bar{\psi}\epsilon})\delta\epsilon_\mu .
\end{split}
  \label{eq:variation_decoupled_fermions}
\end{align}
In Eq.~\eqref{eq:variation_decoupled_fermions} the fermionic and
bosonic fluctuations are now completely decoupled and the part which
depends on the $\epsilon$-scalars can be written as
\begin{align}
\delta^2\Lag_{\Phi\epsilon}=-\frac{1}{2}\begin{pmatrix}
\delta\epsilon_\mu & \delta \Phi
\end{pmatrix}
\begin{pmatrix}
\tilde{\Omega}^{\mu \nu}_{\epsilon} & \tilde{X}^\mu_{\epsilon\text{Ph}} \\
\tilde{X}^\nu_{\text{Ph}\epsilon} & \Delta_{\text{Ph}}
\end{pmatrix}
\begin{pmatrix}
\delta\epsilon_\nu \\ \delta \Phi
\end{pmatrix}\text{,}
\end{align}
where
\begin{align}
\tilde{\Omega}^{\mu \nu}_{\epsilon}&=\tilde{\Lambda}^{\mu\nu}_{\epsilon}-2\tilde{\Xi}^\mu_{\epsilon \psi}\Lambda^{-1}_{\psi}\tilde{\Xi}^\nu_{ \bar{\psi}\epsilon} \text{,} \\
\tilde{X}^\mu_{\epsilon\text{Ph}}&=\tilde{\Xi}^\mu_{\epsilon\Phi}-2\tilde{\Xi}^\mu_{\epsilon \psi}\Lambda^{-1}_{\psi}\Xi_{ \bar{\psi}\Phi}\text{,}  \\
\tilde{X}^\mu_{\text{Ph}\epsilon}&=\tilde{\Xi}^\mu_{\Phi\epsilon}-2\Xi_{\Phi \psi}\Lambda^{-1}_{\psi}\tilde{\Xi}^\mu_{ \bar{\psi}\epsilon}\text{,}  \\
\Delta_{\text{Ph}}&=\Lambda_{\Phi}-2\Xi_{\Phi \psi}\Lambda^{-1}_{\psi}\Xi_{ \bar{\psi}\Phi}\text{.}
\end{align}
The term $\Delta_{\text{Ph}}$ does not depend on the
$\epsilon$-scalars. Performing the path integral over the
$\epsilon$-scalars and the scalars $\Phi_i$ we find the effective
action
\begin{align}
\Gamma=\frac{i}{2}\log \det \begin{pmatrix}
\tilde{\Omega}^{\mu \nu}_{\epsilon} & \tilde{X}^\mu_{\epsilon\text{Ph}} \\
\tilde{X}^\nu_{\text{Ph}\epsilon} & \Delta_{\text{Ph}}
\end{pmatrix}\equiv\frac{i}{2}\log \det Q .
\end{align}
The matrix $Q$ can be brought into a diagonal form by inserting
$U$ and $V$ to the left and to the right of $Q$ and by
choosing
\begin{align}
  U & =\begin{pmatrix}
    \mathds{1} & -\tilde{X}_{\epsilon \text{Ph}}\Delta_{\text{Ph}}^{-1} \\
    0 & \mathds{1}
  \end{pmatrix} , \\
  V & =\begin{pmatrix}
    \mathds{1} & 0 \\
    -\Delta_{\text{Ph}}^{-1}\tilde{X}_{\text{Ph}\epsilon} & \mathds{1}
  \end{pmatrix} .
\end{align}
The resulting effective action reads
\begin{align}
  \Gamma=\frac{i}{2}\log \det\left(\tilde{\Omega}_\epsilon ^{\mu\nu}-\tilde{X}^\mu_{\epsilon\text{Ph}}\Delta_\text{Ph}^{-1}\tilde{X}^\nu_{\text{Ph}\epsilon}\right)+\frac{i}{2}\log\det\Delta_\text{Ph} ,
  \label{finalgamma}
\end{align}
where only the first term depends on the $\epsilon$-scalars.
Substituting the expressions for $\tilde{\Omega}_\epsilon ^{\mu\nu}$,
$\tilde{X}^\mu_{\epsilon\text{Ph}}$, $\Delta_\text{Ph}^{-1}$ and
$\tilde{X}^\nu_{\text{Ph}\epsilon}$ into the first term we find the
$\epsilon$-dependent part
\begin{align}
  \label{gammaeps}
  \Gamma &= \frac{i}{2}\log \det\left(\tilde{\Lambda}^{\mu\nu}_{\epsilon}-2\tilde{\Xi}^\mu_{\epsilon \psi}\Lambda^{-1}_{\psi}\tilde{\Xi}^\nu_{ \bar{\psi}\epsilon}-\tilde{W}^{\mu\nu}\right) + \cdots , \\
  \tilde{W}^{\mu \nu}&=\left(\tilde{\Xi}^\mu_{\epsilon\Phi}-2\tilde{\Xi}^\mu_{\epsilon \psi}\Lambda^{-1}_{\psi}\Xi_{ \bar{\psi}\Phi}\right)
     \left(\Lambda_{\Phi}-2\Xi_{\Phi \psi}\Lambda^{-1}_{\psi}\Xi_{ \bar{\psi}\Phi}\right)^{-1}
     \left(\tilde{\Xi}^\nu_{\Phi\epsilon}-2\Xi_{\Phi \psi}\Lambda^{-1}_{\psi}\tilde{\Xi}^\nu_{ \bar{\psi}\epsilon} \right) .
\end{align}
In a standard EFT calculation Eq.~\eqref{gammaeps} is written as a
trace in momentum space and must be expanded in powers of $p/M$ to
obtain a local action, where $M$ is the mass of the heavy particle to
be integrated out.  In our calculation, however, all 1-loop integrals
get multiplied by $\epsilon$, so only the divergent parts give a
non-zero contribution to $\Gamma$ in the limit $\epsilon\to 0$.  Since
the divergences in a renormalizable gauge theory are local
\cite{Weinberg:1959nj,Zimmermann:1968mu}, we obtain a local action.
By performing a power counting we find that the only terms that yield
divergent momentum integrals are
\begin{align}
  \label{finalgammaeps}
  \Gamma_{\text{div}}&=\frac{i}{2}\log \det\left(\tilde{\Delta}^{\mu\nu}_{\epsilon}-2\tilde{Y}^{\mu\nu}_{\lambda}-2\tilde{Y}^{\mu\nu}_{\psi}+2\tilde{Z}^{\mu \nu}_{\lambda \psi}+2\tilde{Z}^{\mu \nu}_{\psi \lambda}\right) , \\
  \tilde{Y}^{\mu\nu}_\omega&=\tilde{X}^\mu_{\epsilon \omega}\Delta^{-1}_{\omega}\tilde{X}^\nu_{ \bar{\omega}\epsilon} , \\
  \tilde{Z}^{\mu\nu}_{\omega\sigma}&=\tilde{X}^\mu_{\epsilon \omega}\Delta^{-1}_{\omega}X_{\bar{\omega}\sigma}\Delta_{\sigma}^{-1}\tilde{X}^\nu_{\bar{\sigma}\epsilon} .
\end{align}
Using the results of Ref.~\cite{Ellis:2017jns} and the methods described in Ref.~\cite{Zhang:2016pja} we find
the following effective Lagrangian containing all contributions from
integrating out the $\epsilon$-scalars,
\begin{align}
\begin{split}
  16 \pi^2 \epsilon \Lag_\text{reg} =
&-\sum _{i} (m^2_{\epsilon})_{i} (\tilde{X}^\mu _{\epsilon \epsilon \mu})_{ii}
 + \frac{1}{2} \sum_{ij} (\tilde{X}^{\mu}_{\epsilon \epsilon \nu})_{ij} (\tilde{X}^{\nu}_{\epsilon \epsilon \mu})_{ji}  \\ 
&+\sum_{ij} 2^{c_{F_j}} \left[2 m_{\psi j} (\tilde{X}^\mu_{\epsilon \psi})_{ij} (\tilde{X} _{\bar{\psi} \epsilon \mu})_{ji} + (\tilde{X}^\mu_{\epsilon \psi})_{ij} i \hat{D}_\nu \hat{\gamma}^\nu (\tilde{X}_{\bar{\psi} \epsilon \mu})_{ji}\right] \\
&-\sum_{i j k} 2^{c_{F_j}+c_{F_k}-1} (\tilde{X}^\mu_{\epsilon \psi})_{ij} \hat{\gamma} ^\nu (X_{\bar{\psi} \psi})_{jk} \hat{\gamma}_{\nu} (\tilde{X}_{\bar{\psi} \epsilon \mu})_{ki} \\ 
& + \frac{\epsilon}{12} \tr\left[ \hat{G}'_{\mu \nu} \hat{G}'^{\mu \nu} \right] ,
\end{split}
\label{masterformula}
\end{align}
where $\hat{G}'_{\mu\nu}=-ig\hat{G}^a_{\mu\nu}t^a$,
$\hat{G}^a_{\mu\nu} = \hat{\partial}_\mu \hat{A}^a_\nu -
\hat{\partial}_\nu \hat{A}^a_\mu + g f^{abc} \hat{A}_\mu^b
\hat{A}_\nu^c$ and $c_F=0$ for Dirac fermions and $c_F=1$ for Majorana
fermions. All quantities with Lorentz indices appearing in
Eq.~\eqref{masterformula} are still projected onto either $QdS$ or
$Q\epsilon S$. After inserting the respective functional derivatives
into this equation each term on the right hand side will contain a
factor $\epsilon$. One can then divide the equation by $\epsilon$ and
take the limit $\epsilon \to 0$. After this limit has been taken there
is no difference between $d$-dimensional and $4$-dimensional
quantities anymore and the hats can be removed. It should be pointed
out that in Eq.~\eqref{masterformula} the Latin indices contain all
indices (generation, gauge, \ldots\@), except for the Lorentz indices
of the $\epsilon$-scalars. Thus, the sums are to be interpreted as a
trace over all indices with the coefficient given by
Eq.~\eqref{masterformula}. Also, we consider the Majorana spinors
$\lambda$ and $\bar{\lambda}$ to be independent. This convention has
to be followed when calculating quantities like
$X_{\bar{\lambda}\lambda}$ from the Lagrangian of the full model. Furthermore, we
stress that the order of $\hat{D}_\nu$ and $\hat{\gamma}^\nu$ in the
second line matters, whenever $\hat{D}_\nu$ contains chiral
projectors.

In the next Section we apply Eq.~\eqref{masterformula} to reproduce
the general parameter relations given in Ref.~\cite{Martin:1993yx} for a
supersymmetric Lagrangian.  However, we'd like to remark that
Eq.~\eqref{masterformula} is a generalization of the results of
Ref.~\cite{Martin:1993yx}, because it contains terms that correspond
to field renormalizations and tadpoles and can be applied also to
non-supersymmetric models regularized in DRED.

\section{Applications}
\label{sec:applications}

In this section we apply Eq.~\eqref{masterformula} to reproduce the
parameter relations given in Ref.~\cite{Martin:1993yx} for a
supersymmetric Lagrangian with the gauge coupling $g$ corresponding to a simple gauge group, a gaugino mass
parameter $M$, a gaugino--fermion--scalar coupling $\gc$, a Yukawa
coupling $Y^{ijk}$ and a quartic scalar coupling $\lambda^{ij}_{kl}$.
\subsection{Gauge coupling}
\label{sec:gauge_coupling}

The relation between the DRED and the DREG gauge coupling can be
obtained from the last term in Eq.~\eqref{masterformula}, where the limit $\epsilon \to 0$ can be taken immediately,
\begin{align}
  \mathcal{L}_\text{reg,gauge} = \frac{1}{12(16\pi^2)} \tr\left[G'_{\mu \nu} G'^{\mu \nu}\right]
  = -\frac{g^2}{12(16 \pi^2)} C(G) G^a_{\mu \nu} G_a ^{\mu \nu} ,
  \label{eq:Delta_L_gauge}
\end{align}
where $G'_{\mu\nu}=-igG^a_{\mu\nu}t^a$ and $t^a$ are the generators in
the adjoint representation of the gauge group,
$(t^a)^{bc} = -if^{abc}$, and
$C(G) \delta^{ab} = \tr[t^at^b] = f^{acd}f^{bcd}$.  The term in
Eq.~\eqref{eq:Delta_L_gauge} can be absorbed into a finite field
renormalization of the gauge field and a shift in the gauge coupling,
\begin{align}
  (A_{\mu}^a)^\DRED &\to \left(1 - \frac{1}{2}\delta Z_A\right) (A^a_\mu)^\DREG , \\
  g^\DRED &\to g^\DREG - \delta g ,
\end{align}
with
\begin{align}
   \delta Z_A&=\frac{g^2}{3(16\pi^2)} C(G) , \\
  \delta g &= -\frac{1}{2} g \delta Z_A . \label{eq:delta_g_2}
\end{align}
From Eq.~\eqref{eq:delta_g_2} one obtains
\begin{align}
  g^{\DREG} &= g^{\DRED} \left(1 - \frac{g^2}{6(16\pi^2)} C(G) \right) , \label{eq:Delta_g}
\end{align}
which agrees with the result of Ref.~\cite{Martin:1993yx}.

\subsection{Gaugino mass parameter}
\label{sec:gaugino_mass}

We assume that the supersymmetric Lagrangian (regularized in DRED)
contains the kinetic and the soft-breaking gaugino mass term
\begin{align}
  \Lag_{\lambda} = \frac{1}{2} \bar{\lambda}^a \left(i \hat{\slashed{\partial}} - M^\DRED \right) \lambda^a ,
\end{align}
where $\lambda^a$ denotes the gaugino Majorana spinor, transforming in
the adjoint representation of the gauge group.  In addition, there is
an interaction term between the gauginos and the $\epsilon$-scalars,
\begin{align}
  \Lag_{\epsilon\lambda} =
  \frac{g}{2} \bar{\lambda}^b \tilde{\gamma}^\mu \epsilon^a _\mu (t^a)_{bc} \lambda^c ,
\end{align}
with $t^a_{bc} = -i f^{abc}$.  When the $\epsilon$-scalars are
integrated out, the following two terms from the second line of
Eq.~\eqref{masterformula} contribute to the relation between $M^\DRED$
and $M^\DREG$:
\begin{align}
  16\pi^2\epsilon \mathcal{L}_{\text{reg},\lambda} =
   2 \left(2 M \tilde{X}^\mu _{\epsilon \lambda}  \tilde{X}_{ \bar{\lambda} \epsilon \mu} +  \tilde{X}^\mu _{\epsilon \lambda}i \hat{D}_\nu \hat{\gamma}^\nu  \tilde{X}_{ \bar{\lambda} \epsilon \mu}\right) .
\end{align}
The derivatives $ \tilde{X}^\mu_{\epsilon \lambda}$ and $\tilde{X}_{ \bar{\lambda} \epsilon}^\mu$
are obtained from $\Lag_{\epsilon\lambda}$ and read
\begin{align}
  \label{Xepsi}
  (\tilde{X}_{\epsilon \lambda}^{\mu})^a_b &= \frac{g}{2} \bar{\lambda}_c\tilde{\gamma}^\mu (t^a)_{cb} , \\
  \label{Xpsie}
  (\tilde{X}_{ \bar{\lambda} \epsilon}^{\mu})^a_b &= - \frac{g}{2} \tilde{\gamma}^\mu (t^a)_{bc}\lambda_c ,
\end{align}
which yields
\begin{align}
  16\pi^2\epsilon \mathcal{L}_{\text{reg},\lambda} &=
  \frac{g^2}{2} \bar{\lambda}^a \left(
  - \tilde{\gamma}^\mu i\hat{\slashed{\partial}}\tilde{\gamma}_\mu - 2 M \tilde{\gamma}^\mu\tilde{\gamma}_\mu
  \right) (t^b)_{ac}(t^b)_{cd} \lambda^d , \\
  &= \frac{g^2}{2} \epsilon \bar{\lambda}^a \left(
  i\hat{\slashed{\partial}} - 2 M
  \right) C(G) \lambda^a ,
  \label{eq:L_reg_lambda}
\end{align}
with $(t^b)_{ac}(t^b)_{cd} = f^{abc} f^{dbc} = C(G) \delta^{ad}$ and
$\tilde{\gamma}^\mu \tilde{\gamma}_\mu = \epsilon$. After dividing by $\epsilon$ and taking the limit $\epsilon \to 0$ the
terms in Eq.~\eqref{eq:L_reg_lambda} can be absorbed by the finite
field and parameter re-definitions
\begin{align}
  (\lambda^a)^\DRED &\to \left( 1 - \frac{1}{2}\delta Z_\lambda \right) (\lambda^a)^\DREG , \\
  M^\DRED &\to M^\DREG - \delta M ,
\end{align}
with
\begin{align}
  \delta Z_\lambda &= \frac{g^2}{16\pi^2} C(G) , \\
  \delta M &= M\left( 2\frac{g^2}{16\pi^2} C(G) - \delta Z_\lambda \right) .
\end{align}
Thus, the relation between the gaugino mass parameter in DRED and DREG
reads
\begin{align}
  M^\DREG = M^\DRED + \delta M =
  M^\DRED \left( 1 + \frac{g^2}{16\pi^2} C(G) \right) ,
\end{align}
which is in agreement with the result of Ref.~\cite{Martin:1993yx}.

\subsection{Gaugino coupling}
\label{sec:gaugino_coupling}
We consider a supersymmetric and gauge invariant Lagrangian with a
gaugino $\lambda^a$, charged scalars $\phi_i$ and Dirac fermions
$\psi_i$.  The left- and right-handed components of the $\psi_i$ are
assumed to originate from superfields transforming in the (generally
reducible) representation $R$ and its conjugate representation
$\bar{R}$, respectively. The mass eigenstates are obtained from a
diagonalization of a mass matrix by two unitary matrices $L$ and
$R$. The scalar fields $\phi_{1i}$ and $\phi_{2i}$ originate from the
same superfields as the left- and right-handed components of the Dirac
fermions, respectively, and the scalar mass eigenstates are obtained
from the diagonalization of the mass matrix with the unitary matrix
$U$.  The Lagrangian, formulated in terms of the mass eigenstate
fields $\psi_i$, $\phi_{1i}$ and $\phi_{2i}$ then contains the
coupling term
\begin{align}
  \Lag_{\gc} &= \sqrt{2}\gc\Big(\phi^{*} _{1i} U^{\dagger} _{ij}
\bar{\lambda}^{a} (t^{a} _{R})_{jk}L_{km}P_L\psi _{m}-\bar{\psi}_{i} R^T_{ij}P_{L}(t^{a}_{R})_{jk}\lambda^{a} U^{*} _{km} \phi^{*}_{2m}+\text{h.c.}\Big) ,
\label{gauginofermion}
\end{align}
where $t_R^a$ denotes the generator in the representation of the Dirac
fields $\psi_i$. Here the Latin indices run over both flavor indices and gauge group indices.
In DRED supersymmetry ensures that $\gc^\DRED = g$, where $g$
denotes the gauge coupling.  In DREG supersymmetry is explicitly
violated and one has $\gc^\DREG \neq g$.  To the relation between
$\gc^\DRED$ and $\gc^\DREG$ the third line of
Eq.~\eqref{masterformula} contributes, which reads for the considered
case
\begin{align}
  16\pi^2 \epsilon \Lag_{\text{reg},\gc}&=
 -\tilde{X}^\mu _{\epsilon \lambda} \hat{\gamma}^\nu  X_{\bar{\lambda} \psi} \hat{\gamma}_{\nu} \tilde{X} _{ \bar{\psi} \epsilon \mu}
 -\tilde{X}^\mu _{\epsilon \psi} \hat{\gamma}^\nu  X_{\bar{\psi} \lambda} \hat{\gamma}_{\nu}  \tilde{X} _{\bar{\lambda} \epsilon \mu} .
 \label{eq:gaugino_coupling_contrib}
\end{align}
The derivatives $\tilde{X}^\mu_{\epsilon \lambda}$ and
$\tilde{X}^\mu_{\bar{\lambda} \epsilon}$ have been calculated in
Section~\ref{sec:gaugino_mass} already.  The derivatives
$\tilde{X}_{\epsilon \psi}^{\mu}$ and
$\tilde{X}_{ \bar{\psi} \epsilon}^{\mu}$ can be obtained from the Dirac
fermion--$\epsilon$-scalar coupling of the DRED Lagrangian
\begin{align}
  \Lag_{\epsilon\bar{\psi}\psi} = g\epsilon^a _\mu \bar{\psi}_i\tilde{\gamma}^\mu\left( R^T_{ij}(t_R ^a)_{jk} R^{*}_{kl} P_R+L^{\dagger}_{ij}(t_R ^a)_{jk} L_{kl} P_L \right) \psi_l ,
\end{align}
which yields
\begin{align}
  (\tilde{X}_{\epsilon \psi}^{\mu})^a_l &= g\bar{\psi}_i\tilde{\gamma}^\mu (T^a_R)_{il} , \\
  (\tilde{X}_{ \bar{\psi} \epsilon}^{\mu})^a_i &= - g\tilde{\gamma}^\mu (T^a_R)_{il}\psi_l ,
\end{align}
where we have introduced the abbreviation
\begin{align}
  (T^a_R)_{il} = R^T_{ij}(t_R ^a)_{jk} R^{*}_{kl}
  P_R+L^{\dagger}_{ij}(t_R ^a)_{jk} L_{kl} P_L  .
\end{align}
The derivatives $X_{\bar{\lambda} \psi}$ and $X_{\bar{\psi} \lambda}$
can be read off Eq.~\eqref{gauginofermion} and read
\begin{align}
  (X_{\bar{\lambda} \psi})^a_j &= \sqrt{2} \gc\left[\phi^{*}_{1i} (U^{\dagger} t^a_{R} L)_{ij}P_{L}-\phi_{2i} (U^{T}t^a_{R}R^{*})_{ij} P_{R}  \right]\equiv \sqrt{2}\gc A^a_j , \\
\label{Xpsilam}
  (X_{ \bar{\psi} \lambda})^a_i &= \sqrt{2} \gc \left[(L^{\dagger}t^a_{R}U)_{im}\phi_{1m}P_{R}-(R^Tt^a_{R} U^*)_{im}P_{L}\phi^{*}_{2m}\right]\equiv \sqrt{2}\gc B^a_i\text{.}
\end{align}
Inserting all derivatives into Eq.~\eqref{eq:gaugino_coupling_contrib} yields
\begin{align}
  16\pi^2 \epsilon \Lag_{\text{reg},\gc}&=
 \frac{\gc g^2}{\sqrt{2}} \left[
   \bar{\lambda}^i\tilde{\gamma}^\mu (t_G^a)_{ij} \hat{\gamma}^\nu A^j_k\hat{\gamma}_\nu\tilde{\gamma}_\mu(T^a_R)_{kl}\psi_l
 + \bar{\psi}_i\tilde{\gamma}^\mu (T_R^a)_{ij} \hat{\gamma}^\nu B^k_j\hat{\gamma}_\nu\tilde{\gamma}_\mu(t^a_G)_{kl}\lambda_l \right]
\nonumber \\ &=\frac{d}{4}\sqrt{2}\gc g^2\epsilon C(G)\left(\bar{\lambda}^iA^i_l\psi_l+\bar{\psi}_iB^l_i\lambda^l
\right) ,
\label{eq:gaugino_operator}
\end{align}
where we used that
\begin{align}
  \tilde{\gamma}^\mu \tilde{\gamma}_\mu &= \epsilon , \\
  (t^a_G)_{ij}A^j_k(T^a_R)_{kl}&=\frac{1}{2}C(G)A^i_l , \\
  (t^a_G)_{kl}(T^a_R)_{ij}\gamma^\mu B^k_j&=\frac{1}{2}\gamma^\mu C(G)B^l_i ,
\end{align}
and $(t_G^a)_{bc} = -if^{abc}$ being the generators in the adjoint
representation of the gauge group.  In addition to the term on the
r.h.s.\ of Eq.~\eqref{eq:gaugino_operator} finite field
renormalizations of the Dirac fermions and of the gaugino contribute
to the difference between $\gc ^\DRED$ and $\gc ^\DREG$.  The
field renormalization of the gaugino has already been calculated in
Section~\ref{sec:gaugino_mass}. The field renormalization of the
Dirac fermion follows from the second term in the second line of
Eq.~\eqref{masterformula}, which reads
\begin{align}
 16\pi^2\epsilon \Lag_{\text{reg},\bar{\psi}\psi} = - g^2 \bar{\psi}_i\tilde{\gamma}^\mu (T_R^a)_{il} i\hat{\slashed{\partial}}\tilde{\gamma}_\mu (T^a_R)_{lk}\psi_k .
  \label{eq:Lam_reg_psi_psi}
\end{align}
After taking the limit $\epsilon \to 0$ the terms on the r.h.s.\ of Eqs.~\eqref{eq:Lam_reg_psi_psi} and
\eqref{eq:gaugino_operator} can be absorbed by the finite field and
parameter re-definitions
\begin{align}
  \psi_i^\DRED &\to \left( \delta_{ij} - \frac{1}{2} (\delta Z_\psi)_{ij} \right) \psi_j^\DREG , \\
  g_\lambda ^{\DRED} &\to \left( 1 - \delta \gc \right) g_\lambda ^\DREG ,
\end{align}
where
\begin{align}
  (\delta Z_\psi)_{ij} &= \frac{g^2}{16\pi^2} C(r_i)\delta_{ij} , \\
  \delta \gc &= \frac{g^2}{32\pi^2}C(G) - \frac{1}{2} \delta Z_\psi ,
\label{FFR}
\end{align}
and we used $(T^a_R)_{il}(T^a_R)_{lk}=C(r_i) \delta_{ik}$. Here, the $r_i$ are the irreducible components of the representation $R$ and the index $i$ is not summed over. From
Eq.~\eqref{FFR} one obtains the relation
\begin{align}
\label{eq:gauginocorr}
  g_\lambda ^\DREG &= g^\DRED \left(1 + \delta\gc \right)
               = g^\DRED \left(1+\frac{g^2}{32\pi^2}\left[C(G)-C(r_i)\right]\right) ,
\end{align} 
which depends on the irreducible representation in which the chiral
superfield transforms.  The relation \eqref{eq:gauginocorr} agrees
with the result of Ref.~\cite{Martin:1993yx}.

\subsection{Yukawa coupling}

We consider a supersymmetric and gauge invariant Lagrangian with the
superpotential
\begin{align}
  \mathcal{W} = \frac{1}{6} Y_{ijk} \Phi _i\Phi _j \Phi _k ,
\end{align}
where $\Phi_i$ are chiral superfields and $Y_{ijk}$ is the Yukawa
coupling.  Furthermore we assume that the Weyl fermionic components of
the superfields can be arranged into Dirac fermions $\psi_i$. The
left- and right-handed components of these Dirac fermions are assumed
to originate from the diagonalization of a mass matrix using the
unitary matrices $R$ and $L$. The scalar fields $\phi_i$ may originate
from the diagonalization of the scalar mass matrix with the unitary
matrix $U$. The Lagrangian, formulated in terms of the mass eigenstate
fields $\psi_i$ and $\phi_i$, then contains the Yukawa coupling term
\begin{align}
  \mathcal{L}_y = \frac{1}{6}Y_{ijk}U_{il}\phi_l L_{jm}R_{kn}\bar{\psi} _{n}P_L\psi_{m} + \text{h.c.}
  \equiv\frac{1}{6}\Upsilon_{lmn}\phi_l \bar{\psi} _{n}P_L\psi _{m} + \text{h.c.} ,
\label{YukawaLag}
\end{align}
where $\Upsilon_{lmn}=Y_{ijk}U_{il}L_{jm}R_{kn}$.  The DRED to DREG
parameter conversion for the Yukawa coupling receives field
renormalization contributions from $\psi_i$, which
originate from the the second term in the second line of
Eq.~\eqref{masterformula}.  In addition, the term in the third line of
Eq.~\eqref{masterformula} (and its hermitian conjugate) give an
explicit contribution to the Yukawa coupling, which reads
\begin{align}
  16\pi^2\epsilon \Lag_{\text{reg},y} = 
  -\frac{1}{2}\tilde{X}^\mu _{\epsilon \psi} \hat{\gamma} ^\nu X_{\bar{\psi} \psi} \hat{\gamma}_{\nu} \tilde{X} _{ \bar{\psi} \epsilon \mu} .
  \label{fullProduct}
\end{align}
The appearing derivatives read
\begin{align}
  X_{\bar{\psi}_n\psi_m}&=-\frac{1}{6}\Upsilon_{lmn}\phi_l P_L , \\
  (\tilde{X}_{\epsilon \psi}^{\mu})^a_l &= g\bar{\psi}_i\tilde{\gamma}^\mu (T^a_{R_\psi})_{il} , \\
  (\tilde{X}_{\bar{\psi} \epsilon}^{\mu})^a_i &= - g\tilde{\gamma}^\mu (T^a_{R_\psi})_{il}\psi_l ,
\end{align}
and one obtains
\begin{align}
  16\pi^2\epsilon \Lag_{\text{reg},y} =
  -\frac{d}{12}g^2\epsilon\bar{\psi}_i (T^a_{R_\psi})^F_{ij}(\Upsilon_{lkj}\phi_l P_L)(T^a_{R_\psi})_{km}\psi_m ,
\label{firstYukRenorm}
\end{align}
where we have used that $\hat{\gamma}^\nu \hat{\gamma}_\nu = d$,
$\tilde{\gamma}^\mu \tilde{\gamma}_\mu = \epsilon$ and we have defined
\begin{align}
  (T^a_{R_\psi})_{il} &= R^T_{ij}(t_{R_\psi}^a)_{jk} R^{*}_{kl} P_R+L^{\dagger}_{ij}(t_{R_\psi}^a)_{jk} L_{kl} P_L , \\
  (T^a_{R_\psi})^F_{il} &= R^T_{ij}(t_{R_\psi}^a)_{jk} R^{*}_{kl} P_L + L^{\dagger}_{ij}(t_{R_\psi}^a)_{jk} L_{kl} P_R .
\end{align}
The gauge invariance of Eq.~\eqref{YukawaLag} implies
\begin{align}
Y_{lnj}(t^a_{R_\psi})_{mj}=Y_{jnm}(t^a_{R_\phi})_{jl}+Y_{ljm}(t^a_{R_\psi})_{jn}\text{,}
\end{align} 
where $(t^a_{R_\phi})$ are the generators of the representation under
which the scalar fields transform. Using this relation one can simplify the r.h.s.\ of
Eq.~\eqref{firstYukRenorm} by writing
\begin{align}
  (T^a_{R_\psi})^F_{ij}(\Upsilon_{lkj}\phi_l P_L)(T^a_{R_\psi})_{km} =
  \frac{1}{2}\Upsilon_{lmi}\phi_l P_L \big[ C(r_{\psi,m})+C(r_{\psi,i})-C(r_{\phi,l}) \big] ,
\end{align}
which yields
\begin{align}
  16\pi^2\epsilon \Lag_{\text{reg},y} =
  -\frac{d}{24}g^2\epsilon\bar{\psi}_i\Upsilon_{lmi}\phi_l P_L \big[C(r_{\psi,m})+C(r_{\psi,i})-C(r_{\phi,l})\big]\psi_m,
\end{align}
for the irreducible representations $r_{\psi,m}$, $r_{\psi,i}$ and $r_{\phi,l}$. This term and the appearing terms bilinear in the fields $\psi_i$ can be absorbed by the finite field and parameter
re-definitions
\begin{align}
  \psi_i^\DRED &\to \left( 1 - \frac{1}{2} \delta Z_{\psi,i} \right) \psi_i^\DREG , \\
  Y^\DRED &\to \left(1 - \delta Y \right) Y^\DREG ,
\end{align}
with
\begin{align}
  \delta Z_{\psi,i} &= \frac{g^2}{16\pi^2} C(r_{\psi,i}) , \\
  \delta Y_{lmn} &= \frac{g^2}{16\pi^2}\big[C(r_{\psi,m})+C(r_{\psi,n})-C(r_{\phi,l})\big]  - \frac{1}{2} \left( \delta Z_{\psi,m} + \delta Z_{\psi,n} \right),
  \label{eq:delta_Y}
\end{align}
where the limit $\epsilon \to 0$ has been taken.
From Eq.~\eqref{eq:delta_Y} one obtains the relation
\begin{align}
Y_{lmn}^{\DREG}&=Y_{lmn} ^{\DRED} \left( 1 + \delta Y \right) , \\
             &=Y_{lmn} ^{\DRED}\left\{1 + \frac{g^2}{32\pi^2}\big[C(r_{\psi,m})+C(r_{\psi,n})-2C(r_{\phi,l})\big]\right\} ,
               \label{eq:yukawacorr}
\end{align}
which agrees with the result of Ref.~\cite{Martin:1993yx}.

\subsection{Quartic scalar coupling}
\label{sec:quartic_scalar_coupling}

Here we reproduce the known result for the relation between
quartic scalar couplings in DRED and DREG.  We consider a general
gauge invariant (not necessarily supersymmetric) Lagrangian with the
quartic scalar coupling term
\begin{align}
  \Lag_{\lambda}=-\frac{1}{4}\lambda_{ijkl}\varphi^*_i \varphi^*_j \varphi_k \varphi_l .
\end{align}
We assume that the gauge eigenstate fields $\varphi_i$ are rotated
into mass eigenstates $\phi_i$ with a unitary matrix $U$.
The only contribution to the relation between $\lambda^\DRED$ and
$\lambda^\DREG$ originates from the second term in the first line of
Eq.~\eqref{masterformula},
\begin{align}
  16 \pi^2 \epsilon \mathcal{L}_{\text{reg},\lambda} =
  \frac{1}{2} \sum_{i j} (\tilde{X}^{\mu}_{\epsilon \epsilon \nu})_{ij} (\tilde{X}^{\nu}_{\epsilon \epsilon \mu})_{ji} \,.
\end{align}
The derivative $(\tilde{X}^{\nu}_{\epsilon \epsilon \mu})_{ji}$ can be
obtained from the kinetic term of the scalar fields,
$(D_\mu \phi)^\dagger _i (D^\mu \phi)_i$, which contains the coupling
to $\epsilon$-scalars. These couplings are of the form
\begin{align}
  \Lag_{\epsilon\phi} = g^2 \phi^*_i (T^a)_{ij} (T^b)_{jk} \phi_k \epsilon_\mu^a \epsilon_\nu^b \tilde{g}^{\mu \nu}\text{,}
  \label{eq:epsscalarcoupling}
\end{align}
where $(T^a)=U^\dagger t^a U$. From this coupling we find
\begin{align}
 \tilde{X}^{\mu \nu}_{\epsilon \epsilon}=g^2\tilde{g}^{\mu \nu} \phi^*_i \{T^a,T^b\}_{ij} \phi_j
\end{align}
and the contribution to the effective Lagrangian is
\begin{align}
  16\pi^2 \epsilon \Lag_{\text{reg},\lambda} &=\frac{g^4}{2} \tilde{g}^\mu_\nu \phi_i^* \{T^a,T^b\}_{ij} \phi_j \tilde{g}^\nu _\mu \phi^*_k\{T^b,T^a\}_{kl}\phi_l \\
  &=\frac{g^4}{2} \epsilon \varphi_i^* \varphi_k^* \varphi_j \varphi_l \{t^a,t^b\}_{ij} \{t^b,t^a\}_{kl} , \label{eq:Lam_reg_phi4}
\end{align}
where we have used
$\tilde{g}^\mu_\nu \tilde{g}^\nu_\mu = \tilde{g}^\mu_\mu = \epsilon$.  The
term on the r.h.s.\ of Eq.~\eqref{eq:Lam_reg_phi4} can be absorbed by
the parameter re-definition
\begin{align}
  \lambda^\DRED = \lambda^\DREG - \delta \lambda
\end{align}
with
\begin{align}
  \lambda^{\DREG}_{ijkl} &= \lambda^{\DRED}_{ijkl} + \delta\lambda , \\
  &= \lambda^{\DRED}_{ijkl}
  -\frac{g^4}{16\pi^2} \Big(\{t^a,t^b\}_{ik}\{t^b,t^a\}_{jl}+\{t^a,t^b\}_{il}\{t^b,t^a\}_{jk} \Big) .
\label{eq:quarticcorr}
\end{align}
The relation \eqref{eq:quarticcorr} agrees with the result of
Ref.~\cite{Martin:1993yx}.

\subsection{Trilinear, quadratic and tadpole couplings}
\label{sec:tadpole_coupling}

In a supersymmetric gauge theory, renormalized in the \DRbarPrime
scheme, without spontaneous symmetry breaking the quartic scalar
coupling $\lambda_{ijkl}$ is the only coupling from the scalar
potential which receives a non-zero contribution from the
$\epsilon$-scalars \cite{Martin:1993yx}.  In a spontaneously broken
gauge theory, however, there may be additional non-zero contributions
to the trilinear, quadratic and tadpole scalar couplings from the
$\epsilon$-scalars.  These non-zero contributions originate from
replacing the scalar fields $\phi_i$ by non-zero vacuum expectation
values $v_i$ (VEVs) and corresponding perturbations $\eta_i$, as
$\phi_i = v_i + \eta_i$.  Therefore it is expected that the
contribution to the other scalar couplings from the $\epsilon$-scalars
is proportional to the VEVs.  In this section we calculate the
relation of the trilinear, quadratic and tadpole scalar couplings
between DRED and DREG in a general spontaneously broken gauge theory
using the result of Eq.~\eqref{masterformula}.  We consider a theory
with a simple gauge group $G$ that is spontaneously broken by the VEVs
of some real scalar fields $\phi_i$.  The scalar potential in such a
general renormalizable gauge theory reads
\begin{align}
-V(\phi)=\xi_i \phi_i +\frac{1}{2} m^2_{ij} \phi_i \phi_j +\frac{1}{3} h_{ijk}\phi_i \phi_j \phi_k+\frac{1}{4} \lambda_{ijkl} \phi_i \phi_j \phi_k \phi_l,
\end{align}
where all couplings are totally symmetric.  Expanding the scalar
fields around their VEVs as $\phi_i = v_i + \eta_i$ yields the
potential
\begin{align}
\begin{split}
-V(\eta)&=\xi_i v_i+\frac{1}{2} m^2_{ij} v_i v_j +\frac{1}{3} h_{ijk} v_i v_j v_k +\frac{1}{4} \lambda_{ijkl} v_i v_j v_k v_l \\
&\quad +\left(\xi_i+m^2_{ij}v_j+h_{ijk} v_j v_k+\lambda_{ijkl} v_j v_k v_l\right)\eta_i \\
&\quad +\left(\frac{1}{2}m^2_{ij}+h_{ijk}v_k+\frac{3}{2} \lambda_{ijkl} v_k v_l \right) \eta_i \eta_j  \\
&\quad +\left(\frac{1}{3}h_{ijk}+\lambda_{ijkl}v_l\right)\eta_i \eta_j \eta_k+\frac{1}{4}\lambda_{ijkl} \eta_i \eta_j \eta_k \eta_l \,.
\end{split}
\label{eq:tree-potential}
\end{align}
In order for a minimum to be attained at $\eta_i=0 \, \forall i$ the
following conditions must be satisfied
\begin{align}
\xi_i+m^2_{ij}v_j+h_{ijk} v_j v_k+\lambda_{ijkl} v_j v_k v_l=0 \qquad \forall i \text{.}
\end{align}
When integrating out the $\epsilon$-scalars one obtains corrections to
the potential \eqref{eq:tree-potential} from the first line of
Eq.~\eqref{masterformula},
\begin{align}
  16 \pi^2 \epsilon \Lag_{\text{reg},\eta} &=
  -\sum_{a} (m^2_{\varepsilon})_{a} (\tilde{X}^\mu_{\varepsilon \varepsilon \mu})_{aa}
  +\frac{1}{2} \sum_{ab} (\tilde{X}^{\mu}_{\varepsilon \varepsilon \nu})_{ab} (\tilde{X}^{\nu}_{\varepsilon \varepsilon \mu})_{ba} \,.
  \label{eq:masterformula_scalar}
\end{align}
The derivatives $(\tilde{X}^{\mu\nu}_{\varepsilon \varepsilon})_{ab}$
in Eq.~\eqref{eq:masterformula_scalar} are to be taken with respect to
the $\epsilon$-scalar mass eigenstates, denoted by $\varepsilon_a$, and
$(m^2_{\varepsilon})_{a}$ are the corresponding mass eigenvalues.  The
derivatives can be calculated from the interaction Lagrangian between
the $\epsilon$-scalars and the scalar fields $\eta_i$, which reads
\begin{align}
\Lag_{\epsilon\eta}&=\frac{g^2}{2} \epsilon_\mu^a \epsilon_\nu^b \tilde{g}^{\mu \nu} (T^a)_{ij} (T^b)_{ik}
  \left( v_j v_k + 2 \eta_j v_k + \eta_j \eta_k \right),
\label{eq:Lag_eps_eta}
\end{align}
where $T^a=-it^a$ are real, antisymmetric matrices and $t^a$ are the
generators of the representation under which the $\eta_i$ transform.
The first term in the parentheses of \eqref{eq:Lag_eps_eta}
contributes to the mass matrix of $\epsilon$-scalars, which reads
\begin{align}
  (m^2_\varepsilon)_{ab} = m^2_\epsilon \delta_{ab}+g^2(T^a)_{ij} v_j (T^b)_{ik} v_k\text{.}
\end{align}
Since this matrix is symmetric it can be diagonalized by an orthogonal
matrix $O$ such that
$O_{ab} (m^2_\varepsilon)_{bc} O_{dc} = (m^2_\varepsilon)_a
\delta_{ad}$. The
corresponding mass eigenstates $\varepsilon_a$ are then given by
$\varepsilon_a = O_{ab} \epsilon_b$ and the interaction Lagrangian in
terms of $\varepsilon$-scalar mass eigenstates becomes
\begin{align}
  \Lag_{\varepsilon\eta}&=\frac{g^2}{2} \varepsilon_\mu^a \varepsilon_\nu^b \tilde{g}^{\mu \nu}
  (T_O^a)_{ij} (T_O^b)_{ik} \left(2 \eta_j v_k + \eta_j \eta_k \right),
  \label{eq:Lag_eps_eta_2}
\end{align}
where $T_O^a=O_{ab}T^b$ and we have omitted the $\varepsilon$-scalar
mass term.  From Eq.~\eqref{eq:Lag_eps_eta_2} one obtains the
derivatives
\begin{align}
  (\tilde{X}^{\mu \nu}_{\varepsilon \varepsilon})^{ab} &= -\frac{g^2}{2}\tilde{g}^{\mu \nu}
    \{T_O^a,T_O^b\}_{kj} \left(2v_k \eta_j+\eta_k \eta_j\right) , \\
(\tilde{X}^\mu_{\varepsilon \varepsilon \mu})^{aa} &= -\frac{g^2}{2}\epsilon
    \{T^a,T^a\}_{kj} \left(2v_k\eta_j + \eta_k\eta_j\right) ,
\end{align}
and the contribution from the $\epsilon$-scalars becomes
\begin{align}
\begin{split}
  16 \pi^2 \epsilon \Lag_{\text{reg},\eta} &=
  (m^2_{\varepsilon})_{a} \frac{g^2}{2}\epsilon \{T^a,T^a\}_{kj} \left(2v_k\eta_j + \eta_k\eta_j\right) \\
  &\quad + \frac{g^4}{8}\epsilon  \{T^a,T^b\}_{kj}\{T^a,T^b\}_{lm} \left(4v_k v_l \eta_j \eta_m + 4v_k \eta_l \eta_j \eta_m 
     + \eta_k \eta_l \eta_j \eta_m\right) .
\end{split}
\label{eq:Lag_reg_scalar}
\end{align}
From Eq.~\eqref{eq:Lag_reg_scalar} one can see that for
$v_i = 0\ \forall i$ there would only be a contribution to the
quadratic and to the quartic scalar coupling, as was pointed out in
Ref.~\cite{Martin:1993yx}.  However, when $v_i \ne 0$ these two
contributions also get distributed to other terms in the scalar
potential.  The new scalar potential including the contribution from
the $\epsilon$-scalars becomes
\begin{align}
\begin{split}
-V(\eta)&=\xi_i v_i+\frac{1}{2} m^2_{ij} v_i v_j +\frac{1}{3} h_{ijk} v_i v_j v_k +\frac{1}{4} \lambda_{ijkl} v_i v_j v_k v_l \\
&\quad +\left(\xi_i+m^2_{ij}v_j+h_{ijk} v_j v_k+\lambda_{ijkl} v_j v_k v_l+\frac{1}{16\pi^2}v_k \A_{ki}\right)\eta_i \\
&\quad +\left(\frac{1}{2}m^2_{ij}+h_{ijk}v_k+\frac{3}{2} \lambda_{ijkl} v_k v_l+\frac{1}{2(16\pi^2)}\A_{ij}+\frac{1}{2(16\pi^2)}v_l v_k \B_{kilj}\right) \eta_i \eta_j \\
&\quad +\left(\frac{1}{3}h_{ijk}+\lambda_{ijkl}v_l+\frac{1}{2(16\pi^2)}v_l\B_{lijk}\right)\eta_i \eta_j \eta_k \\
&\quad+\left(\frac{1}{4}\lambda_{ijkl}+\frac{1}{8(16\pi^2)}\B_{ijkl}\right) \eta_i \eta_j \eta_k \eta_l\text{,}
\end{split}
\label{eq:loop-potential}
\end{align}
where we have introduced the abbreviations
\begin{align}
\A_{ij}&\equiv g^2 (m^2_\varepsilon)_a \{T^a,T^a\}_{ij} = 2 g^2 (m^2_\varepsilon)_a (T^a T^a)_{ij}, \\
\B_{ijkl}&\equiv g^4 \{T^a,T^b\}_{ij}\{T^a,T^b\}_{kl},
\end{align} 
and all repeated indices are summed over.
In Eq.~\eqref{eq:loop-potential} all parameters are still defined in
\DRED.  The 1-loop terms on the r.h.s.\ of
Eq.~\eqref{eq:loop-potential} can be absorbed by the parameter
re-definitions
\begin{align}
  p^{\DRED}=p^{\DREG} - \delta p,
\end{align}
where $p\in\{\xi,m^2,h,\lambda,v\}$.  Note, that
$\eta^\DRED = \eta^\DREG$, because there is no contribution to the
field renormalization of scalar fields from Eq.~\eqref{masterformula}.
By demanding that the potential written in terms of \DREG parameters
takes the same form as in Eq.~\eqref{eq:tree-potential} we obtain the
following set of equations relating the shifts to the finite loop
corrections from the $\epsilon$-scalars
\begin{align}
&\delta \lambda_{ijkl}=\frac{1}{2(16\pi^2)}\B_{(ijkl)}, \\
&\frac{1}{3} \delta h_{ijk}+\delta \lambda_{ijkl} v_l+\lambda_{ijkl}\delta v_l=\frac{1}{2(16\pi^2)}v_l\B_{l(ijk)}, \\
&\frac{1}{2}\delta m^2_{ij}+\delta h_{ijk}v_k+h_{ijk}\delta v_k+\frac{3}{2}\delta \lambda_{ijkl} v_k v_l+3\lambda_{ijkl}\delta v_k v_l=\frac{\A_{(ij)}}{2(16\pi^2)}+\frac{v_k v_l\B_{k(i|l|j)}}{2(16\pi^2)}, \\
&\delta \xi_i+\delta m^2_{ij} v_j+m^2_{ij}\delta v_j+\delta h_{ijk} v_j v_k+2h_{ijk}\delta v_j v_k+\delta \lambda_{ijkl} v_j v_l v_k+3\lambda_{ijkl}\delta v_j v_k v_l=\frac{v_k\A_{ki}}{16\pi^2},
\end{align}
where
$T_{(i_1 i_2\dots i_n)}\equiv\frac{1}{n!}\sum_{\sigma \in
  S_n}T_{\sigma (i_1) \sigma (i_2) \dots \sigma (i_n)}$ and $S_n$ is
the symmetric group over $n$ symbols and $\B_{k(i|l|j)}=\frac{1}{2}\sum_{\sigma \in
  S_2}B_{k \sigma (i) l \sigma (j)}$.
This set of equations is equivalent to
\begin{align}
  \label{eq:lambdashift}
  \delta \lambda_{ijkl} &= \frac{1}{2(16\pi^2)}\B_{(ijkl)}, \\
  \label{eq:hshift}
  \frac{1}{3} \delta h_{ijk}+\lambda_{ijkl}\delta v_l &= \frac{v_l}{2(16\pi^2)}\left[\B_{l(ijk)}-\B_{(ijkl)}\right], \\
  \label{eq:mshift}
  \frac{1}{2}\delta m^2_{ij}+h_{ijk}\delta v_k &= \frac{\A_{ij}}{2(16\pi^2)}
     + \frac{v_k v_l}{2(16\pi^2)} \left[
       \B_{k(i|l|j)} - 3\B_{l(ijk)} + \frac{3}{2}\B_{(ijkl)} \right], \\
  \label{eq:xishift}
  \delta \xi_i + m^2_{ij}\delta v_j &= \frac{v_j v_k v_l}{(16\pi^2)} \left[
    -\B_{k(i|l|j)}+\frac{3}{2}\B_{l(ijk)} - \frac{1}{2} \B_{(ijkl)} \right],
\end{align}
where we have used that $\A_{(ij)} = \A_{ij}$, because
$\A_{ij}=\A_{ji}$.  Eq.~\eqref{eq:lambdashift} is equivalent to the
result obtained in Section~\ref{sec:quartic_scalar_coupling} for
complex scalar fields.  The Eqs.~\eqref{eq:hshift}--\eqref{eq:xishift}
can be simplified further by using the fact that the shifts of the
vacuum expectation values $\delta v_i$ can be related to the shifts
$\delta Z_\eta$ of the scalar fields and corresponding (auxiliary)
background fields $\delta \hat{Z}_{\eta}$ as
\cite{Sperling:2013eva,Sperling:2013xqa}
\begin{align}
  \delta v_i = \frac{1}{2} \left( \delta Z_\eta + \delta \hat{Z}_\eta \right)_{ij} v_j \,.
\end{align}
%
%
As pointed out in Refs.~\cite{Sperling:2013eva,Sperling:2013xqa},
neither $\delta Z_\eta$ nor $\delta \hat{Z}_\eta$ receive contributions from
$\epsilon$-scalars, which implies
\begin{align}
  \delta v_i = 0 \qquad \Leftrightarrow \qquad v_i^\DREG = v_i^\DRED \,.
  \label{eq:delta_v}
\end{align}
This allows us to derive the following relations for the trilinear,
quadratic and tadpole scalar couplings between DREG and DRED,
\begin{align}
  \label{eq:hshift_simp}
  h_{ijk}^\DREG &= h_{ijk}^\DRED + \frac{3v_l}{2(16\pi^2)}\left[\B_{l(ijk)}-\B_{(ijkl)}\right], \\
  \label{eq:mshift_simp}
  (m^2_{ij})^\DREG &= (m^2_{ij})^\DRED + \frac{\A_{ij}}{(16\pi^2)}
     + \frac{v_k v_l}{(16\pi^2)} \left[
       \B_{k(i|l|j)} - 3\B_{l(ijk)} + \frac{3}{2}\B_{(ijkl)} \right], \\
  \label{eq:xishift_simp}
  \xi_i^\DREG &= \xi_i^\DRED + \frac{v_j v_k v_l}{(16\pi^2)} \left[-
    \B_{k(i|l|j)}+\frac{3}{2}\B_{l(ijk)} - \frac{1}{2} \B_{(ijkl)} \right] .
\end{align}
The relations \eqref{eq:delta_v}--\eqref{eq:xishift_simp} represent a
generalization of the known results of Ref.~\cite{Martin:1993yx} for a
spontaneously broken gauge theory with non-zero VEVs.  In the limit
$v_i \to 0$, which was used in Ref.~\cite{Martin:1993yx}, one obtains
\begin{align}
  \label{eq:hshift_v_zero}
  h_{ijk}^\DREG &= h_{ijk}^\DRED \,, \\
  \label{eq:mshift_v_zero}
  (m^2_{ij})^\DREG &= (m^2_{ij})^\DRED - \frac{2 g^2}{16\pi^2} m^2_\epsilon C(r_i) \delta_{ij} \,, \\
  \label{eq:xishift_v_zero}
  \xi_i^\DREG &= \xi_i^\DRED \,,
\end{align}
with $(T^aT^a)_{ij} = i^2 (t^a t^a)_{ij} = - C(r_i) \delta_{ij}$.  The
$m^2_\epsilon$-dependence in Eq.~\eqref{eq:mshift_v_zero} can be
removed by shifting the $m^2_{ij}$ parameters as described in
Ref.~\cite{Jack:1994rk}, which is equivalent to transforming from the \DRbar into
the \DRbarPrime scheme as
\begin{align}
  (m^2_{ij})^\DRED = (m^2_{ij})^{\DRED'} + \frac{2 g^2}{16\pi^2} m^2_\epsilon C(r_i) \delta_{ij} \,.
\end{align}
In the \DRbarPrime scheme one therefore obtains in the limit
$v_i \to 0$,
\begin{align}
  \label{eq:shifts_v_zero_DRbarPrime}
  h_{ijk}^\DREG &= h_{ijk}^{\DRED'} \,, &
  (m^2_{ij})^\DREG &= (m^2_{ij})^{\DRED'} \,, &
  \xi_i^\DREG &= \xi_i^{\DRED'} \,,
\end{align}
which is the known result from Ref.~\cite{Martin:1993yx}.

\section{Conclusions}
\label{sec:conclusion}

The universal one-loop effective action (UOLEA) is a very elegant tool
to fully automate the derivation of the large set of effective
Lagrangians of a given UV model with heavy particles.  To date,
however, only part of the UOLEA is known and only in dimensional
regularization (DREG).  Due to this restriction, the known part cannot
be applied to supersymmetric UV models, regularized in dimensional
reduction (DRED), with non-supersymmetric effective theories that are
regularized in DREG.

In this paper we have extended the UOLEA by generic 1-loop operators
which represent a translation between DRED and DREG.  These operators
allow for an application of the UOLEA to supersymmetric UV models with
non-supersymmetric EFTs.  As the UOLEA itself, our derived generic
operators are well suited to be implemented into generic spectrum
generators to fully automate the derivation of non-supersymmetric
EFTs.

We have performed the calculation of the effective operators in the
language of effective field theories, close to the formulation of the
UOLEA.  In our case the field to be integrated out is the unphysical
$\epsilon$-scalar, which occurs in DRED when the $4$-dimensional gauge
field is split into a $d$- and an $\epsilon$-dimensional part.  The
resulting effective operators can be absorbed by a re-definition of
the fields and parameters, leading to the well-known parameter
translations in supersymmetric models of Ref.~\cite{Martin:1993yx}.

In our calculation we have assumed that the UV theory is
renormalizable and gauge invariant, but not necessarily
supersymmetric.  Within these restrictions our result is generic and
contains even terms corresponding to finite field
renormalizations.  Furthermore, our result has an explicit
$m^2_\epsilon$ dependence, which can be removed by shifting the
squared mass parameters of the scalar fields appropriately as
described in Ref.~\cite{Jack:1994rk}.

Finally we have applied our derived effective operators to various UV
theories for illustration and to prove that the known results of
Ref.~\cite{Martin:1993yx} can be reproduced.  Furthermore, we have
derived relations for scalar cubic, quadratic and tadpole couplings in
a general renormalizable gauge theory with spontaneous symmetry
breaking, complementing the results of Ref.~\cite{Martin:1993yx}.

\acknowledgments

We kindly thank Jérémie Quevillon for helpful discussions regarding
the UOLEA.  A.V.\ is grateful to Robert Harlander for insights into
the $\epsilon$-scalar formalism.  A.V.\ and B.S.\ acknowledge support
by the DFG Research Unit \textit{New Physics at the LHC} (FOR2239).

\clearpage
\appendix
\section{Consistency of shifts for Majorana fermions}
\label{sec:Appendix}

We here show that the shifts \eqref{shift1} and \eqref{shift2} can be
performed consistently for Majorana fermions.  For Majorana fermions
there is only one degree of freedom and so it is necessary that the
shifted fields $\delta\bar{\lambda}'$ and $\delta\lambda'$ are related
as $\delta \bar{\lambda}' = \delta \lambda'^\dagger \gamma^0$.  That this is in fact the case follows from the
Hermiticity of the Lagrangian. Consider the terms containing Majorana
fermions in the original Lagrangian
\begin{align}
\label{majolag}
\mathcal{L}_\lambda=\bar{\lambda} F \lambda+\epsilon^a_\mu \bar{\lambda} \tilde{G}^\mu_a \lambda + \phi \bar{\lambda} H \psi + \phi \bar{\psi} I  \lambda+ \phi \bar{\lambda} J \lambda\text{,}
\end{align}
where $F$, $G$, $H$, $I$ and $J$ are independent of the fields. Taking the Hermitian conjugate and using that $\mathcal{L}_\lambda^\dagger=\mathcal{L}_\lambda$ we find
\begin{align}
\mathcal{L}_\lambda=\bar{\lambda} \gamma^0 F^\dagger \gamma^0 \lambda+\epsilon^a_\mu \bar{\lambda} \gamma^0 (\tilde{G}^\mu_a)^\dagger \gamma^0 \lambda + \phi \bar{\lambda} \gamma^0 I^\dagger \gamma^0 \psi + \phi \bar{\psi} \gamma^0 H^\dagger \gamma^0 \lambda\text{,}
\end{align}
which yields the relations
\begin{align}
\label{Frel}
F&=\gamma^0 F^\dagger \gamma^0\text{,} \\
\label{Grel}
\tilde{G}_a^\mu&=\gamma^0 (\tilde{G}_a^\mu)^\dagger \gamma^0\text{,} \\ 
\label{Hrel}
H&=\gamma^0 I^\dagger \gamma^0\text{,} \\
\label{Irel} 
I&=\gamma^0 H^\dagger \gamma^0\text{,} \\
\label{Jrel}
J&=\gamma^0 J^\dagger \gamma^0\text{.}
\end{align}
We can also relate $F$, $G$, $H$, $I$ and $J$ to the quantities appearing in the second variation of the Lagrangian by noting that
\begin{align}
\begin{split}
\delta^2\mathcal{L}_\lambda&=\delta \bar{\lambda} (F+\epsilon^a_\mu \tilde{G}_a^\mu+\phi J) \delta \lambda + \delta \bar{\lambda} \tilde{G}_a^\mu\lambda \delta \epsilon^a_\mu+\delta \epsilon^a_\mu \bar{\lambda} \tilde{G}_a^\mu \delta \lambda+\delta \bar{\lambda} (H\psi+J\lambda) \delta \phi \\
&\quad+\delta \bar{\lambda} H\phi \delta \psi+\delta \bar{\psi} I\phi \delta \lambda+\delta \phi (\bar{\lambda}J+\bar{\psi}I)\delta \lambda + \cdots\text{,}
\end{split}
\end{align}
where the extra terms indicated by the ellipsis do not include any
variation of $\lambda$ or $\bar{\lambda}$. Comparing this to
\eqref{fullsecondvar} one obtains
\begin{align}
\Delta_\lambda&=(F+\epsilon^a_\mu \tilde{G}_a^\mu+\phi J)\text{,} \\
\tilde{X}^\mu_{\epsilon \lambda}&= \bar{\lambda} \tilde{G}_a^\mu\text{,} \\
\tilde{X}^\mu_{\bar{\lambda}\epsilon}&=-\tilde{G}_a^\mu\lambda\text{,} \\
X_{\phi \lambda}&=\bar{\lambda}J+\bar{\psi}I\text{,} \\
X_{\bar{\lambda}\phi}&=-(H\psi+J\lambda)\text{,} \\
X_{\bar{\psi} \lambda}&=I\phi\text{,} \\
X_{\bar{\lambda}\psi}&=H\phi \text{.}
\end{align}
From these relations and \eqref{Frel}--\eqref{Jrel} it follows that 
\begin{align}
\label{dagger1}
(\Delta^\dagger_\lambda)^{-1}&=\gamma^0 \Delta_\lambda^{-1} \gamma^0\text{,} \\
(\tilde{X}^\mu_{\bar{\lambda \epsilon}})^\dagger&=-\tilde{X}^\mu_{\epsilon \lambda}\gamma^0\text{,} \\
X^\dagger_{\bar{\lambda} \phi}&=-X_{\phi \lambda} \gamma^0\text{,} \\ 
\label{daggerdone}
X^\dagger _{\bar{\lambda} \psi}&=\gamma^0 X_{\bar{\psi} \lambda}\gamma^0 \text{.} 
\end{align}
Calculating the Dirac adjoint of the shift \eqref{shift1} we obtain
\begin{align}
\begin{split}
\delta \lambda'^{\dagger} \gamma^0&=\delta \bar{\lambda}-\left(\tilde{X}^{\mu \dagger} \delta \epsilon_\mu+X^\dagger_{\bar{\lambda} \phi}\delta \phi-\delta \psi^\dagger X^\dagger_{\bar{\lambda}\psi}\right)(\Delta_\lambda^{-1})^\dagger \gamma^0
\\ &=\delta \bar{\lambda}-\left(-\tilde{X}^\mu_{\epsilon \lambda}\gamma^0 \delta \epsilon_\mu-X_{\phi \lambda}\gamma^0\delta \phi-\delta \psi^\dagger \gamma^0 X_{\bar{\psi}\lambda}\gamma^0\right)\gamma^0(\Delta_\lambda^{-1})\gamma^0\gamma^0 \\
&=\delta \bar{\lambda}+\left(\tilde{X}^\mu_{\epsilon \lambda}\delta \epsilon_\mu+X_{\phi \lambda}\delta \phi +\delta \bar{\psi} X_{\bar{\psi}\lambda}\right)\Delta_\lambda^{-1} \\
&=\delta \bar{\lambda}'\text{,}
\end{split}
\end{align}
where we have used Eqs.~\eqref{dagger1}--\eqref{daggerdone} in the
second line and the definition \eqref{shift2} in the last line.  We
conclude that the shifts \eqref{shift1}--\eqref{shift2} are consistent
with the required property for Majorana fermions,
$\delta\bar{\lambda}' = \delta\lambda'^{\dagger}\gamma^0$.

\bibliographystyle{JHEP}
\bibliography{paper.bib}

\end{document}